\documentclass[aps,prb,twocolumn,groupedaddress,floats,showpacs]{revtex4}
\usepackage{latexsym}
\usepackage{dcolumn}
\usepackage[dvips]{graphicx}
\usepackage{amssymb,amsmath,bm}
\usepackage{graphics}
\usepackage{epsf}

\begin{document}
\title{Density of states of disordered graphene}
\author{Ben Yu-Kuang Hu$^{1,2}$, E. H. Hwang$^1$,
and S. Das  Sarma$^1$} \affiliation{$^1$Condensed Matter Theory
Center, Department of
        Physics, University of Maryland, College Park, MD 20742-4111}
\affiliation{$^2$Department of Physics, The University of Akron,
Akron, OH 44325-4001}
\date{\today}
\begin{abstract}

We calculate the average single particle density of states in
graphene with disorder due to impurity potentials. For unscreened
short-ranged impurities, we use the non-self-consistent and
self-consistent Born and $T$-matrix approximations to obtain the
self-energy. Among these, only the self-consistent $T$-matrix
approximation gives a non-zero density of states at the Dirac
point.  The density of states at the Dirac point is non-analytic
in the impurity potential.  For screened short-ranged and charged
long-range impurity potentials, the density of states near the
Dirac point typically increases in the presence of impurities,
compared to that of the pure system.

\end{abstract}
 \maketitle

\section{introduction}

The recent experimental realization of a single layer of carbon
atoms arranged in a honey-comb lattice has prompted much
excitement and activity in both the experimental
and theoretical
physics communities \cite{review,neto}.  Carriers in graphene
(both electrons and holes) have a linear bare kinetic energy
dispersion spectra around the $K$ and $K'$ points (the ``Dirac
points") of the Brillouin zone.  The ability of experimentalists
to tune the chemical potential to lie above or below the Dirac
point energy (by application of voltages to gates in close
proximity to the graphene sheets) allows the carriers to be
changed from electrons to holes in the same sample.  This sets
graphene apart from other two dimensional (2D) carrier systems
that have a parabolic dispersion relation, and typically have only
one set of carriers, i.e. either electrons or holes. Another
unique electronic property, the absence of back-scattering, has
led to the speculation that carrier mobilities of 2D graphene
monolayers (certainly at room temperature, but also at low
temperature) could be made to be much higher than any other
field-effect type device, suggesting great potential both for
graphene to be the successor to Si-MOSFET
(metal-oxide-semicondcutor field effect transistor) devices and
for the discovery of new phenomena that normally accompanies any
significant increase in carrier mobility
\cite{kn:hwang2006c,kn:ando2006,bolotin,geim}. It is therefore of
considerable fundamental and technological interest to understand
the electronic properties of graphene \cite{neto}.

Graphene samples that are currently being fabricated are far from
pure, based on the relatively low electronic mobilities compared
to epitaxially grown modulation-doped two-dimensional electron
gases (2DEGs) such as GaAs-AlGaAs quantum wells.  It is therefore
important to understand the effects of disorder on the properties
of graphene. Disorder manifests itself in the finite lifetimes of
electronic eigenstates of the pure system.  In the presence of
scattering from an impurity potential that is not diagonal in the
these eigenstates, the lifetime scattering rate of the eigenstate,
$\gamma$, is non-zero and can be measured experimentally by
fitting the line-shape of the low-field Shubnikov-de Haas (SdH)
oscillations. \cite{kn:Harrang85,kn:Lukyanchuk06}  The effect of
disorder scattering on the SdH line shape is equivalent to
increasing the sample temperature and one can therefore measure
this Dingle temperature ($T_D$) and relate it to the
single-particle lifetime through $\hbar\gamma=2\pi k_B T_D$.  To
avoid potential confusion, we mention that the lifetime damping
rate $\gamma$ discussed in this paper is {\em not} equal to the
{\sl transport} scattering rate which governs the electrical
conductivity.  The lifetime damping rate is the measure of the
rate at which particles scatters out of an eigenstate, whereas the
transport scattering rate is a measure of the rate of {\sl current
decay} due to scattering out of an eigenstate.  In normal 2DEGs,
the transport scattering time can be much larger than the impurity
induced lifetime $\hbar/2\gamma$, particularly in high mobility
modulation doped 2D systems where the charged impurities are
placed very far from the 2DEG\cite{kn:Harrang85,kn:Dassarma85}.
Recently, the issue of transport scattering time versus impurity
scattering lifetime in graphene has been
discussed.\cite{hwang_single}

The single-particle level broadening due to the impurity potential
changes many of the physical properties of the system including
the electronic density of  states.\cite{kn:Dassarma81,kn:ando1982}
The electronic density of states is an important property which
directly affects many experimentally measurable quantities such as
the electrical conductivity, thermoelectric effects, and
differential conductivity in tunneling experiments between
graphene and scanning-tunneling microscope tips or other electron
gases.  Changes in the density of states also modify the electron
screening,\cite{kn:Ando_s82} which is an important factor in the
determination of various properties of graphene.  It is therefore
imperative to take into account the effects of disorder on the
density of states, particularly since disorder is quite strong in
currently available graphene samples.

In the present work, we present calculations of the average
density of states of disordered graphene.  This problem has been
investigated using various models and techniques, both analytical
and
numerical.\cite{hu1984,fradkin,peres2006,pereira,skrypnyk,dora2008,wu2008,lherbier2008}
We take into account scattering effects from long-range and short
range impurity potentials. We consider both unscreened and
screened short-ranged and screened charged impurities, using the
Born approximation.  In addition, for unscreened short-ranged
(USR) impurities, we go beyond the Born approximation and include
self-consistent effects.

There is another class of disorder in graphene called
``off-diagonal" or ``random gauge potential" disorder, in which
the hopping matrix elements of the electrons in the underlying
honeycomb lattice are random.   In this paper we do not consider
in this type of disorder, which can result from height
fluctuations (ripples) in the graphene sheet and lead to
qualitatively different results from the ones presented in this
paper.\cite{hu1984,dora2008,morita1997,guinea2008}

The rest of  the paper is organized as follows: In
Sec.~\ref{sec:approx}, we describe the approximation schemes that
we use.  Secs.~\ref{sec:USR} and \ref{sec:screened} deal with
unscreened short range impurities and screened short-range/charged
impurities, respectively. In Sec.~\ref{sec:comparison}, we compare
our results to those from other workers, and we conclude in
Sec.~\ref{sec:conclusion}.

\section{Approximations for the self-energy}
\label{sec:approx}

The single-particle density of states for a translationally
invariant 2DEG is given by\cite{rickmah}
\begin{align}
D(E) &= -\frac{g}\pi \sum_{\lambda} \int \frac{d\bm k}{(2\pi)^2}\
{{\rm Im}}[G_{\lambda}(\bm k,E)] \label{eq:DoS}
\end{align}
where $g$ is the degeneracy factor (for graphene $g=4$ due to
valley and spin degeneracies) $\lambda$ is the band index, $G$ is
the retarded Green's function, and the $\bm k$-integration is over
a single valley, which we assume to be a circle of radius $\vert
\bm k\vert = k_c$.  $G$ expressed in terms of the retarded
self-energy $\Sigma_\lambda(\bm k,\omega)$ is
\begin{equation}
G_\lambda(\bm k,E) = [E - E_{\bm k,\lambda} - \Sigma_\lambda(\bm
k,E) + i\eta]^{-1}.\label{eq:G}
\end{equation}
where $E_{\bm k,\lambda}$ is the bare band energy of the state
$\vert \bm k\lambda\rangle$ and $\eta$ is an infinitesimally small
positive number. (In this paper, the Green's functions and
self-energies are all assumed to be retarded.)
Eqs.~(\ref{eq:DoS}) and (\ref{eq:G}) show that if $E_{\bm
k,\lambda}$ and $\Sigma_\lambda(\bm k,E)$ are known, the density
of states can be obtained in principle from
\begin{widetext}
\begin{equation}
D(E) = \frac{g}\pi \sum_{\lambda} \int \frac{d\bm k}{(2\pi)^2}\
\frac{-{{\rm Im}}[\Sigma_{\lambda}(\bm k,E)]+\eta}{(E - E_{\bm
    k,\lambda} - {\rm Re}[\Sigma_\lambda(\bm
  k,E)])^2+({\rm Im}[\Sigma_\lambda(\bm k,E)]-\eta)^2}.
\label{eq:DoS1}
\end{equation}
\end{widetext}


For pure graphene systems, $\Sigma(\bm k,E) = 0$ (excluding
electron--electron and electron--phonon interactions, which are
not considered here), and hence ${\rm Im}[G_\lambda(\bm k,E)] =
-\pi \delta(E-E_{\bm
  k\lambda})$.  Close to the Dirac points (which we choose to the the
zero of energy), the dispersion for graphene is (we use $\hbar =1$
throughout this paper)
\begin{equation}
E_{\bm k,\lambda} = \lambda v_F k,\label{eq:Dirac_dispersion}
\end{equation}
where $\lambda = +1$ and $-1$ for the conduction and valence
bands, respectively, $k = \vert\bm k\vert$ is the wavevector with
respect to the Dirac point, and $v_F$ is the Fermi velocity of
graphene. Performing the $\bm k$-integration in Eq.~(\ref{eq:DoS})
for the pure graphene case gives
\begin{equation}
D_0(E) = \frac{g}{2\pi} \frac{\vert E\vert}{v_F^2}\, \theta(\vert
E\vert - E_c),\label{eq:DoSpure}
\end{equation}
where $E_c =  v_F k_c$ is the band energy cut-off.

The average density of states for a disordered 2DEG can be
obtained by averaging the Green's function over impurity
configurations.  The averaging procedure gives a non-zero $\Sigma$
which, in general, cannot be evaluated exactly. Various
approximation schemes for $\Sigma$ have therefore been developed,
four of which are described below.

\begin{figure}
\epsfysize=1.7in \epsffile{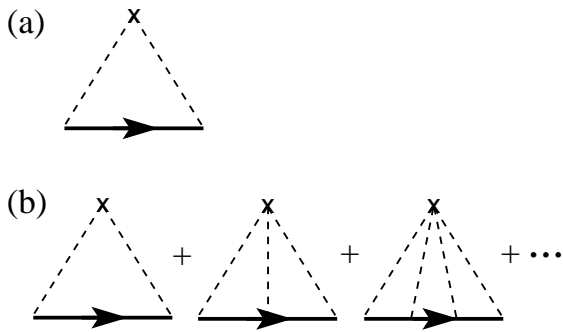} \vspace{0.90cm}
\caption{Feynman diagrams for the (a) Born and (b) $T$-matrix
  approximations for the self-energy.  The ``x", dotted line and line
  with arrow signify the impurity, impurity potential, and Green's
  function, respectively.  The Green's functions are either bare or
  self-consistent.}
\label{fig1}
\end{figure}

{\sl Born Approximation ---} In the Born approximation, the
self-energy is given by the Feynman diagram shown in Fig.
1(a)\cite{rickmah}, and the expression for the self-energy is
\begin{eqnarray}
\Sigma_{B,\lambda}(\bm k,E) = n_i \int \frac{d\bm k'}{(2\pi)^2}\
\vert U(\bm k - \bm
k^\prime)\vert^2\; \nonumber \\
\times  \sum_{\lambda'} G_{0,\lambda'}(\bm
k^\prime,E)\;F_{\lambda\lambda'}(\bm k,\bm k') \label{eq:Sigma_B}
\end{eqnarray}
where $n_i$ is the impurity density, $U(\bm q)$ is the Fourier
transform of the impurity potential, $G_0$ is the bare Green's
function and $F_{\lambda\lambda'}(\bm k,\bm k')$ is square of the
the overlap function between the part of the wavefunctions of
$\vert \bm k\lambda \rangle$ and $\vert \bm
k^\prime\lambda'\rangle$ that are periodic with the lattice (here
$\lambda,\lambda'$ are band indices). For graphene states near the
Dirac point,
\begin{equation}
F_{\lambda\lambda'}(\bm k,\bm k') =
\frac12(1+\lambda\lambda'\cos\theta_{\bm k\bm
k^\prime}),\label{eq:F}
\end{equation} where $\theta_{\bm k\bm k^\prime}$
is the angle between $\bm k$ and $\bm k'$.

{\sl Self-Consistent Born Approximation ---}  The Feynman diagram
for this self-energy is the same as Fig. 1(a), except that the
bare Green's function is replaced by the full one.  Consequently,
the expression for $\Sigma_{\rm SB}$ is the same as in
Eq.~(\ref{eq:Sigma_B}), except with $G_0$ replaced by $G$.

{\sl $T$-matrix Approximation ---}  The $T$-matrix approximation
is equivalent to the summation of Feynman diagrams shown Fig.
1(b).   The expression for $\Sigma_{T}$ is the same as in
Eq.~(\ref{eq:Sigma_B}), except with $U$ replaced by $T$, the
$T$-matrix for an individual impurity.

{\sl Self-consistent $T$-matrix Approximation ---}  The Feynman
diagram for this approximation is the same as in the $T$-matrix
approximation, except that the bare Green's functions are replaced
by full ones. The expression for $\Sigma_{{\rm S}T}$ is the same
as in Eq.~(\ref{eq:Sigma_B}), except with $U$ replaced by $T$, and
$G_0$ replaced by $G$.

When the potentials for the impurities are not all identical (for
example, in the case where there is a distribution of distances of
charged impurities from the graphene sheet), one averages $\vert
U(\bm k-\bm k')\vert^2$ or $\vert T(\bm k-\bm k')\vert^2$ over the
impurities.

\section{Unscreened short-ranged (USR) disorder}
\label{sec:USR}

In the present context, short-ranged impurities are impurities
which result from localized structural defects in the honeycomb
lattice, which are roughly on the length scale of the lattice
constant. In this case, it is acceptable to approximate $U(\bm q)
= U_0$, a real constant, for intravalley scattering processes.  In
this paper, we ignore the intervalley processes.  (However, we
note that if the matrix element joining intervalley states is
constant, inclusion of intervalley scattering in our calculations
is not difficult.)  This simplification allows us to obtain some
analytic expressions for the self-energies in the approximation
schemes mentioned above.


\subsection{Self-energy for USR disorder}

For USR disorder, the four approximation schemes we use give
self-energies that are independent of $\lambda$ and $k$.

\subsubsection{Born approximation}

The self-energy for graphene with USR scatterers in the Born
approximation, using $U(\bm q) = U_0$,
Eqs.~(\ref{eq:Dirac_dispersion}) and (\ref{eq:F}) in
Eq.~(\ref{eq:Sigma_B}), is
\begin{subequations}
\begin{align}
\Sigma^{\rm (usr)}_{B}(E) &=
\tilde\gamma_B H_{0}(E+i\eta);\label{eq:Sigma_B1}\\
\tilde\gamma_B &= \frac{n_i U_0^2}{2v_F^2};\label{eq:tildegammaB}\\
H_{0}(\zeta) &= 2 v_F^2\sum_{\lambda'}\int \frac{d\bm
  k'}{(2\pi)^2}\ G_{0,\lambda'}(\bm k',\zeta) F_{\lambda\lambda'}(\bm
k,\bm k')\nonumber\\
&=\int_0^{E_c} \frac{dE'}{2\pi} \left(\frac{E'}{\zeta - E'}
  + \frac{E'}{\zeta+E'}\right)\nonumber\\
&=-\frac{\zeta}{2\pi} \ln\left(1 - \frac{E_c^2}{\zeta^2}\right).
\label{eq:H_0}
\end{align}
\end{subequations}
As a function of complex $\zeta$, $H_0(\zeta)$ is real and
positive (negative) along the real axis from $E_c$ to $\infty$
($-E_c$ to $-\infty$).  Furthermore, it has a branch cut in on the
real axis of $\zeta$ in between $-E_c$ and $E_c$, so that for
$-E_c < E \ \mbox{(real)} < E_c$,
\begin{equation}
H_0(E\pm i\eta) = -(2\pi)^{-1}\left(E\ln\left\vert
\frac{E_c^2}{E^2}-
    1\right\vert \pm i \pi \vert E \vert\right).
\label{eq:H_0E}
\end{equation}
Fig.~2 shows $H_0(E+i\eta)$.

For real $E$ and $\vert E\vert \ll E_c$,
\begin{equation}
\Sigma^{\rm (usr)}_{B}(E) \approx - \frac{\tilde\gamma_B}2
\left[\frac{2E}\pi\,\ln\left\vert\frac{E}{E_c}\right\vert + i
\vert
  E\vert\right].
\end{equation}
The Born approximation damping rate for state $\vert\bm
k\lambda\rangle$ is $\gamma_B(k) = -2{\rm Im}[\Sigma_B^{\rm
(usr)}(E_{\bm k,\lambda})] =\tilde\gamma_B\,v_F k$.

\begin{figure}
\includegraphics[scale=0.4]{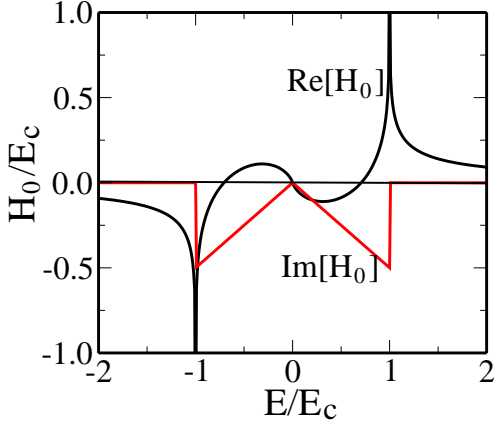}
\caption{Real and imaginary parts of $H_0(E+i\eta) =
\Sigma_B^{\rm
    (usr)}(E)/\tilde\gamma_B$ [see Eq.~(\ref{eq:H_0E})], where
  $\Sigma_B^{\rm (usr)}(E)$ is the self-energy for USR impurities in
  the Born approximation.}
\label{fig2}
\end{figure}


\subsubsection{Self-consistent Born approximation}

The self-energy $\Sigma_{\rm SB}^{\rm (usr)}$ for unscreened
short-ranged scatterers in the self-consistent Born approximation
is given by the self-consistent equation
\begin{subequations}
\begin{align}
\Sigma_{\rm SB}^{\rm (usr)}(E) &=
\tilde\gamma_B\,H_{\rm SB}(E + i\eta);\\
H_{\rm SB}(\zeta)
&=-\frac{1}{2\pi}  \left [ \zeta -\Sigma_{\rm SB}^{\rm
  (usr)}(\zeta)\right ]  \ln\left[1 - \frac{E_c^2}{ [\zeta-\Sigma_{\rm
      SB}^{\rm (usr)}(\zeta) ]^2}\right] \nonumber \\
&= H_0\left[ \zeta-\Sigma_{\rm
    SB}^{\rm (usr)}(\zeta)\right ],\label{eq:Sigma_SB}
\end{align}
\end{subequations}
This shows that if $\vert \Sigma_{\rm SB}^{\rm (usr)}(E) \vert \ll
\vert E \vert$, then $\Sigma_{SB}^{\rm (usr)}(E)\approx
\Sigma_{B}^{\rm (usr)}(E)$ (except possibly around $E = \pm E_c$,
which is usually not experimentally relevant).


\subsubsection{$T$-matrix approximation}

In general the impurity averaged $T$-matrix for a potential $U(\bm
q)$ is
\begin{widetext}
\begin{equation}
T_{\bm k_0\lambda_0,\bm k\lambda}(E) = n_i\sum_{n=1}^\infty
\left[\left(\prod_{i=1}^n  \sum_{\lambda_i=\pm 1}\int \frac{d\bm
      k_i}{(2\pi)^2} U(\bm k_{i-1}-\bm k_i)\,G_{0,\lambda_i}(\bm
    k_i,E)\,F_{\lambda_{i-1},\lambda_i}(\bm k_{i-1},\bm k_i)\right)
  U(\bm k_i-\bm k)\, F_{\lambda_n,\lambda}(\bm k_n,\bm
  k)\right].\label{eq:Tmatrix}
\end{equation}
\end{widetext}
In this approximation, the self energy is
\begin{equation}
\Sigma_{T,\lambda}(\bm k,E) = T_{\bm k\lambda,\bm k\lambda}(E).
\end{equation}
If $U(\bm q) = U_0$, a constant, the term in the square
parentheses in Eq.~(\ref{eq:Tmatrix}) is $U_0^{n+1}
H_0^n(E+i\eta)/(2 v_F^2)^n$.  The sum then gives
\begin{equation}
\Sigma^{\rm (usr)}_T(E) = \frac{\tilde\gamma_B
  H_0(E+i\eta)}{\displaystyle 1- \frac{U_0}{2 v_F^2}
  H_0(E+i\eta)}.\label{eq:Sigma_T}
\end{equation}

\subsubsection{Self-consistent $T$-matrix approximation}

As in self-consistent Born approximation, the $H_0(E)$ in
Eq.~(\ref{eq:Sigma_T}) is replaced by the self-consistent $H_{{\rm
    S}T}(E)$, giving
\begin{subequations}
\begin{align}
\Sigma^{\rm (usr)}_{\mathrm ST}(E) &= \frac{\tilde\gamma_B H_{{\rm
      S}T}(E+i\eta)}{\displaystyle 1- \frac{U_0}{2 v_F^2} H_{{\rm
      S}T}(E+i\eta)};\label{eq:Sigma_ST}\\
H_{{\rm S}T}(\zeta) &= H_0\left(\zeta-\Sigma_{{\rm S}T}^{\rm
(usr)}(\zeta)\right).
\end{align}
\end{subequations}
As in the self-consistent Born approximation, this shows that if
$\vert \Sigma_{{\rm S}T}^{ \rm (usr)}(E) \vert \ll \vert E \vert$,
then $\Sigma_{{\rm  S}T}^{\rm (usr)}(E)\approx \Sigma_{T}^{\rm
(usr)}(E)$ (except possibly around $E = \pm E_c$).


\subsection{Density of states for $\bm k$-independent $\Sigma$}

When the self-energy is $\bm k$- and $\lambda$-independent, the
density of states for graphene can be calculated analytically
from Eq.~(\ref{eq:DoS1}).  The result is
\begin{widetext}
\begin{align}
D(E) &= \frac{g_s g_v}{2\pi^2  v_F^2}\Biggl\{\frac{\Gamma}2
\ln\left(\frac{(E_c^2+\Omega^2 + \Gamma^2)^2 -
    4E_c^2\Omega^2}{(\Omega^2
    +\Gamma^2)^2}\right)+\Omega\left[\tan^{-1}\left(\frac{E_c -
      \Omega}\Gamma\right)
  -\tan^{-1}\left(\frac{E_c+\Omega}\Gamma\right) + 2
  \tan^{-1}\left(\frac\Omega\Gamma\right)\right]\Biggr\},\label{eq:DoS2}
\end{align}
\end{widetext}
where $\Gamma(E) = -{\rm Im}[\Sigma(E)]$ and $\Omega(E) = E - {\rm
Re}\![\Sigma(E)]$.

\subsection{Density of states at the Dirac point}

There has been considerable interest in the minimum DC electrical
conductivity of disordered graphene as the Fermi energy moves
through the Dirac point.\cite{kn:novoselov2005,tan2007}  There is
still no consensus on whether the minimum conductivity is a
universal value or not.  Since the electrical conductivity is
directly proportional to the density of states at the Fermi
energy, it is important to be able to determine the density of
states at the Dirac point of disordered graphene.  Because the
minimum conductivity is non-zero as the Fermi energy passes
through the Dirac point, the density of states should be nonzero.

Eq.~(\ref{eq:DoS1}) shows that if $\Sigma(E)\rightarrow 0$ when
$E\rightarrow 0$, then the density of states at the Dirac point
$D(0) = 0$.  (Note that we have not included the term that is
first order in the impurity potential in our self-energy.  Since
this first-order term merely rigidly shifts the band by an amount
$n_i U_0$, ignoring this term is equivalent to shifting the zero
of the energy by $-n_i U_0$, and hence the Dirac point is still at
$E=0$.)
A non-zero density of states at the Dirac point depends on a
non-zero ${\rm Im} [\Sigma(E=0)]$.   Since $H_0(E\rightarrow 0) =
0$, it is clear from Eqs.~(\ref{eq:Sigma_B1}) and
(\ref{eq:Sigma_T}) that the Born and $T$-matrix approximations
give zero density of states at the Dirac point.

For the self-consistent Born approximation,
Eq.~(\ref{eq:Sigma_SB}) can be rewritten as
\begin{align}
\Sigma_{\rm SB,\lambda}^{\rm (usr)}(E) &= E \left[1 +
  \frac{2\pi}{\tilde\gamma_B\ln\left(1 - \frac{E_c^2}{(E-\Sigma_{\rm
          SB,\lambda}^{\rm (usr)}(E))^2}\right)}\right]^{-1},
\end{align}
which shows that $\Sigma_{\rm SB,\lambda}^{\rm (usr)}(E\rightarrow
0) = 0$, and therefore the self-consistent Born approximation also
gives $D(0) = 0$.

In the case of the self-consistent $T$-matrix approximation,
re-writing Eq~(\ref{eq:Sigma_ST}) as
\begin{equation}
\Sigma_{\mathrm ST}^{\rm (usr)}(E) + n_i U_0 = \frac{n_i U_0}{1 -
  H_0(E-\Sigma_{\mathrm ST}^{\rm (usr)}(E)) U_0/(2 v_F^2)},
\end{equation}
setting $E=0$, using Eq.~(\ref{eq:H_0}) and taking the imaginary
parts of both sides of this equation gives
\begin{align}
{\rm Im}[\Sigma(0)] \equiv\Gamma(0) &= {\rm Im} \left[\frac{n_i
U_0}{\displaystyle 1-i\frac{U_0\Gamma(0)}{2
      v_F^2}\ln\left(1+\frac{E_c^2}{\Gamma(0)^2}\right)}\right].
\label{eq:Gamma0}
\end{align}

In the weak scattering limit, when $\tilde\gamma_B \ll 1$, the
imaginary term in the denominator of Eq.~(\ref{eq:Gamma0}) is much
less than one (we check for self-consistency later), and this
gives
\begin{align}
 \Gamma_{w}(0) &\approx {\rm Im}\left[i \frac{n_i\Gamma_w(0)
     U_0^2}{  v_F^2}
   \ln\left(\frac{E_c}{\Gamma_{w}(0)}\right)\right]
= 2\tilde\gamma_B\Gamma_w(0)
\ln\left(\frac{E_c}{\Gamma_{w}(0)}\right),
\end{align}
which implies that
\begin{equation}
\Gamma_{w}(0) = E_c
\exp\left(-\frac1{2\tilde\gamma_B}\right).\label{eq:Gamma0result}
\end{equation}
Note that the result is non-analytic in $U_0$.  Inserting
Eq.~(\ref{eq:Gamma0result}) into the imaginary part of the
denominator of Eq.~(\ref{eq:Gamma0}) (which we had assumed to be
much smaller than 1 in magnitude) gives the self-consistent
criterion $\exp(-1/2\tilde\gamma_B) \ll n_i U_0/E_c$ for the
validity of Eq.~(\ref{eq:Gamma0result}).   Substituting this into
Eq.~(\ref{eq:DoS2}) gives an average density of states at the
Dirac point for weak scattering of approximately
\begin{equation} \rho_{w}(0) = \frac{g_sg_v E_c}{\pi^2 n_i
U_0^2} \exp\left(-\frac{1}{2\tilde\gamma_B}\right).
\label{eq:DoSweak}
\end{equation}
Similar results to Eq.~(\ref{eq:DoSweak}) have been
reported\cite{fradkin,dora2008,ostrovsky2006} in studies of
disordered systems of fermions with linear dispersions using other
methods.


We mention that our calculation of the graphene density of states
at the Dirac point should only be considered as demonstrative
since electron-electron interaction effects are crucial
\cite{dassarma_ee} at the Dirac point, and the undoped graphene
system is not a simple Fermi liquid at the Dirac point.

\section{Screened short-ranged and charged impurities}
\label{sec:screened}

Free carriers will move to screen a bare impurity potential
$V_{\rm
  ei}(q)$,  resulting in a screened interaction $U(q) = V_{\rm
  ei}(q)/\varepsilon(q)$, where $\varepsilon(q)$ is the static
dielectric function.  The $\varepsilon(q)$ results in an
$q$-dependent effective electron-impurity potential $U$, even in
the case of short-ranged ($q$-independent) bare impurity
potentials.  This makes the calculations much more involved than
in the USR case. Therefore, in this paper, we limit our
investigation of $q$-dependent screened potentials to the level of
the Born approximation.

For the dielectric function $\varepsilon(q)$, we use the random
phase approximation (RPA) for dielectric function appropriate for
graphene, given by
$\varepsilon(q) = 1 - V_c(q) \Pi_0(q)$
where $V_c(q) = 2\pi e^2/(\kappa q)$ is the two-dimensional
Fourier transform of the Coulomb potential ($\kappa$ is the
dielectric constant of the surrounding material), and $\Pi_0(q)$
is the static irreducible RPA polarizability for
graphene.\cite{kn:Hwang2006b} We use $V_{\rm ei}(q) = V_c(q)$ for
charged impurities, and $V_{\rm ei}(q) = U_0$, a constant, for
short-range point defect scatterers.

We first look at the density of states at the Fermi surface; {\em
  i.e.}, at energy $E = \lambda   k_F v_F$.  To obtain this, we
calculate the single-particle lifetime damping rate $\gamma$ in
the Born approximation, which is given by
\begin{eqnarray}
\gamma_\lambda(k) & = & -2{\rm Im}
[\Sigma_{\lambda}(k,E_{k\lambda})] \nonumber \\
& = &\frac{n_i}{2\pi } \frac{k}{v_F}\int_0^{\pi}d\theta\ \frac
{\langle\left |{V_{ei}(q)}\right |^2\rangle}{\varepsilon(q)^2}
(1+\cos\theta), \label{scat_s}
\end{eqnarray}
where $q  = 2k\sin(\theta/2)$.  Then, assuming that ${\rm Im}
[\Sigma_{\lambda}(k,E)]$ is relatively constant for $k$ close to
$k_F$, we substitute $\frac12\gamma_\lambda(k_F)$ for $-{\rm Im}
[\Sigma_\lambda(\bm k,E_F)]$ into Eq.~(\ref{eq:DoS1}), which gives
Eq.~(\ref{eq:DoS2}) with $\Gamma = \frac12 \gamma_\lambda(k_F)$.

We assume that the charged or neutral impurities are distributed
completely at random on the surface of the insulating substrate on
which the graphene layer lies, the areal density for the charged
and neutral impurities is $n_{ic}$ and $n_{i\delta}$,
respectively, and the density of carriers in the graphene layer is
$n = k_F^2/\pi$, where $k_F$ is the Fermi wavevector relative to
the Dirac point. (This relationship between $n$ and $k_F$ takes
into account the spin and valley degeneracy $g_s = 2$ and $g_v =
2$.) We use the RPA screening function at
$T=0$,\cite{kn:Hwang2006b} to obtain the effective impurity
potential. The key dimensionless parameter that quantifies the
screening strength is $r_s = e^2/(\kappa  v_F)$, which is
corresponding to the interaction strength parameter of a normal 2D
system ({\em i.e.} the ratio of potential energy to kinetic
energy). The Born approximation lifetime damping rates for
screened charged-impurities $\gamma_c$ and $\delta$-correlated
neutral impurities $\gamma_\delta$ at $k_F$ are
\begin{subequations}
\begin{align}
\gamma_{c}(k_F) &=  \frac{n_{ic} E_F}{4n}\; I_{c}(2r_s)\label{gamma0c};\\
\gamma_\delta(k_F) &= \frac{2 E_F\tilde\gamma_B}{\pi}\;
I_{\delta}(2r_s)\label{gamma0delta}.
\end{align}
\label{eq:gamma}
\end{subequations}
In these equations,
\begin{subequations}
\begin{align}
I_{c}(x) &= x - \frac{\pi x^2}{2}  + x^3\,f(x);\\
I_{\delta}(x) &= \frac{\pi}4 + 3x\left(1 - \frac{\pi x}2\right) +
x(3x^2 - 2)f(x),
\end{align}
\end{subequations}
where
\begin{align}
f(x) &= \left \{
 \begin{array}{cl} \frac{1}{\sqrt{1-x^2}} \ln\left[\frac{1+\sqrt{1-x^2}}{x}
   \right ]
                  & \mbox{for  $x < 1$} ;\\
                   1
                  & \mbox{for $x =1$}; \\
                   \frac{1}{\sqrt{x^2-1}}\cos^{-1}\frac{1}{x}
                  & \mbox{for $x > 1$}. \end{array}
\right. \label{fx}
\end{align}

\begin{figure}
\includegraphics[scale=0.5]{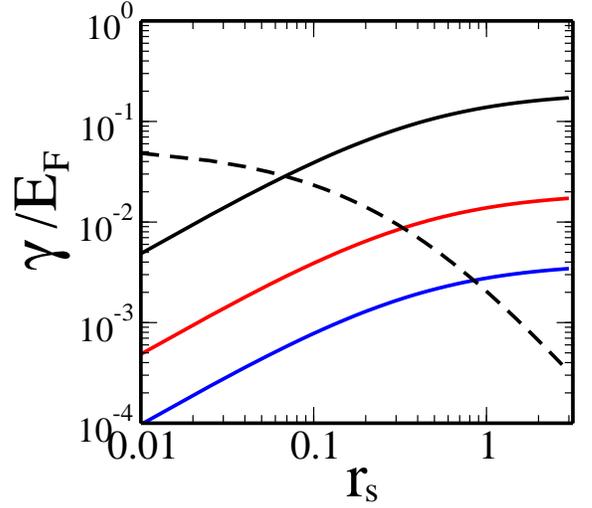}
\caption{Calculated damping rates scaled by Fermi energy
$\gamma/E_F$ as a function of $r_s$.  Graphene on a SiO$_2$ (air)
substrate has an $r_s \approx 0.7$ (2). Solid lines indicate
damping rates ($\gamma_c$) due to charged impurities with an
impurity density $n_{ic} = 10^{11}{\rm cm}^{-2}$ for different
electron densities $n = 1$, 10, 50$\times 10^{11}$ cm$^{-2}$ (from
top to bottom), respectively. Dashed line indicate the damping
rate ($\gamma_{\delta}$) due to short-ranged impurity with
impurity density $n_{i\delta}=10^{11}$ cm$^{-2}$ and potential
strength $U_0=$1 KeV \AA$^2$, which correspond to
$\tilde{\gamma}_B = 0.11$. Note $\gamma_{\delta}/E_F$ is
independent on the electron density.} \label{damping}
\end{figure}

In Fig. \ref{damping} we show the calculated damping rates scaled
by $\vert E_F\vert/  =  k_F v_F$ as a function the interaction
parameter $r_s$. For $r_s\ll 1$ we have ${\gamma_c}/{E_F} \approx
{n_{ic} r_s}/{2n}$ and ${\gamma_{\delta}}/{E_F}  \approx
\tilde{\gamma}_B/2$. For $r_s \gg 1$ we have ${\gamma_c}/{E_F}
\approx \pi n_{ic}
  /{16n}$ and
${\gamma_{\delta}}/{E_F}  \approx  \tilde{\gamma}_B/(2\pi r_s)$.
Thus, for small (large) $r_s$ the damping rate due to the
short-ranged impurity dominates over that due to the long-ranged
charged impurity. On the other hand, since $\gamma_c(k_F) \propto
k_F^{-1} \propto n^{-\frac12}$ and $\gamma_{\delta}(k_F) \propto
k_F \propto n^{\frac12}$ [Eq.~(\ref{eq:gamma})], in the low (high)
carrier density limit the lifetime damping of single particle
states at the Fermi surface is dominated by charged impurity
(short-ranged impurity) scattering.  The crossover takes place
around a density
\begin{equation}
n_{\rm cross} = \frac{n_{ic}}{n_{i\delta}} \frac{\pi
  v_F^2}{4U_0^2} \frac{I_{c}(2r_s)} {I_{\delta}(2r_s)}.
\end{equation}


Using $\Gamma = \gamma_\lambda(k_F)/2$ in Eq.~(\ref{eq:DoS2})
gives (assuming $ E_F$, $\Gamma \ll E_c$)
\begin{eqnarray}
D(E_F) \approx D_0(E_F)\left[ \frac12 + \frac1\pi
  \tan^{-1}\left(\frac{\vert E_F\vert}{\Gamma}\right ) \right . \nonumber \\
\left . +  \frac{\Gamma}{2\pi \vert E_F\vert}
  \ln\left(\frac{E_c^2}{E_F^2+\Gamma^2}\right)\right].
\end{eqnarray}
For $\Gamma/\vert E_F\vert \ll 1$, this gives
\begin{equation}
D(E_F) \approx D_0(E_F)\left\{ 1 +
\frac{1}{\pi}\frac{\Gamma}{\vert E_F\vert} \left[ \ln
  \left(\frac{E_c}{\vert E_F\vert}\right) -1
\right] \right\}, \label{eq:dosdelta}
\end{equation}
and for $\vert E_F\vert/\Gamma \ll 1$
\begin{equation}
D(E_F) \approx D_0(E_F)\left [\frac{1}{2} + \frac{\vert
    E_F\vert}{\pi\Gamma} \right ] + \frac{g_sg_v}{2\pi^2}
\frac{\Gamma}{ v_F^2}\ln\left(\frac{E_c}{\Gamma}\right).
\label{eq:doscharge}
\end{equation}

We can apply Eq. (\ref{eq:dosdelta}) for short-ranged impurity
scattering and for charged impurity scattering in high carrier
density limits, and Eq. (\ref{eq:doscharge}) for charged impurity
scattering in low density limits. Taking the limit $\vert
E_F\vert\rightarrow 0$ in Eq.~(\ref{eq:doscharge}), it appears
that for the case of screened Charged impurities, we obtain a
finite density of states at the Dirac point with the Born
approximation (since $\Gamma \propto \gamma_c(k_F) \propto
k_F^{-1}$).  However, recall that in deriving
Eqs.~(\ref{eq:dosdelta}) and (\ref{eq:doscharge}), we have assumed
that $\Sigma(k,E_F)$ is constant with respect to $k$, which is not
necessarily the case at the Dirac point.

\begin{figure}
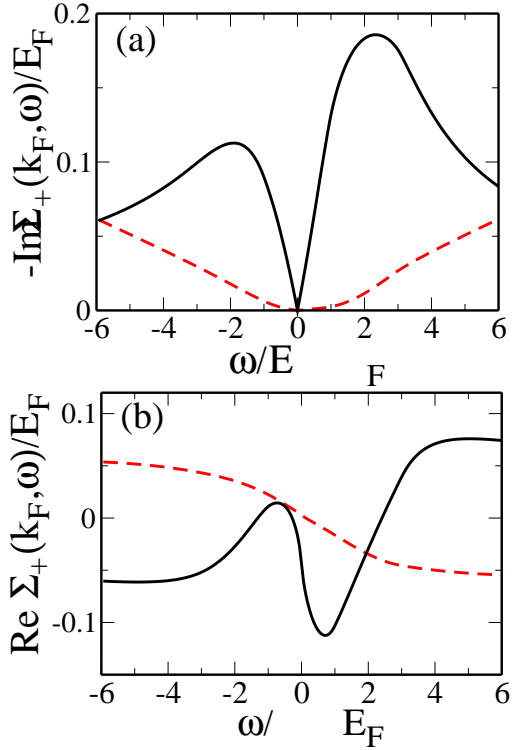

\includegraphics[scale=0.4]{fig4a.eps}
\includegraphics[scale=0.4]{fig4b.eps}
\caption{(a) Imaginary and (b) real parts of self energy of a
  disordered graphene at
$k=k_F$ for screened Coulomb scattering potential (solid lines)
and for screened neutral short-ranged scattering potential (dashed
lines).} \label{self}
\end{figure}

In general, the damping rate (or the imaginary part of the self
energy) is a function of energy and wave vector rather than a
constant. From Eq.~(\ref{eq:Sigma_B}), we calculated the
self-energy of disordered graphene.  Fig.~\ref{self} we show the
self-energy of a conduction band electron ($\lambda=+1$) for both
screened Coulomb scattering potential and screened neutral
short-ranged potential. For Coulomb scatterers we use the impurity
density $n_{ic} = 10^{12}$  cm$^{-2}$,  and for neutral
short-ranged scatterers the impurity density $n_{i\delta}=10^{11}$
cm$^{-2}$ and potential strength $U_0=$1 KeV \AA$^2$. The
self-energies in the valence band ($\lambda =-1$) are related to
the self-energy in the conduction band by
$\rm{Re}\Sigma_+(k,\omega) = -\rm{Re}\Sigma_-(k,-\omega)$ and
$\rm{Im}\Sigma_+(k,\omega) = \rm{Im}\Sigma_-(k,-\omega)$. As
$\omega \rightarrow 0$, $-{\rm Im}\Sigma_{\lambda}(k_F,\omega)
\rightarrow |\omega|$, and ${\rm Re}\Sigma_{\lambda}(k_F,\omega)
\rightarrow \omega \ln |\omega| $ for both scattering potentials.
However, for large value of $\vert\omega\vert$ the asymptotic
behaviors are different, that is, as $\vert\omega\vert \rightarrow
\infty$ $-{\rm Im}\Sigma(k_F,\omega) \propto |\omega|^{-1}$ for
Coulomb scattering potential and $-{\rm Im}\Sigma(k_F,\omega)
\propto |\omega|$ for short-ranged potential. Note that by using
only the (non-self-consistent) Born approximation in this section,
we assume weak scattering and ignoring multiple-scattering events
in calculated $\Sigma(k,\omega)$. Therefore, the results are
unreliable in the strong disorder limit ({\em i.e.}, when
$\Sigma(k,\omega)$ is modified significantly from its lowest-order
form).

\begin{figure}
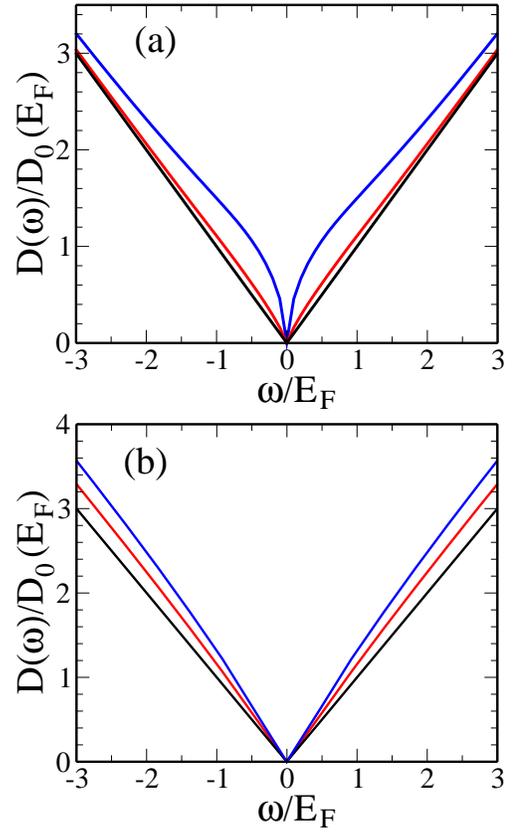

\includegraphics[scale=0.4]{fig5a.eps}
\includegraphics[scale=0.4]{fig5b.eps}
\caption{The density of states in the presence of impurity (a) for
screened Coulomb potential and (b) for screened short-ranged
potential. In (a) we use the charged impurity densities $n_{ic} =
0$, 1, $5\times 10^{12}$ cm$^{-2}$ (from bottom to top), and in
(b) the short-ranged impurity density $n_{id} = 0$, 0.5, $1\times
10^{12}$ cm$^{-2}$ (from bottom to top) and potential strength
$U_0 =$1 KeV \AA$^2$. } \label{dos_f}
\end{figure}

In Fig. \ref{dos_f} the density of states in the presence of
impurity is shown for different impurity densities. In Fig.
\ref{dos_f}(a) we show the density of states for Coulomb impurity
potential with impurity densities, $n_{ic} = 0$, $10^{12}$
cm$^{-2}$, $5\times 10^{12}$ cm$^{-2}$ (from bottom to top), and
in Fig. \ref{dos_f}(b) we show the density of states for neutral
short-ranged impurity potential with densities, $n_{id} = 0$,
$5\times 10^{11}$ cm$^{-2}$, $10^{12}$ cm$^{-2}$ (from bottom to
top) and potential strength $U_0 =$1 KeV \AA$^2$. The calculated
density of states is normalized by $D_0(E_F) =
(g_sg_v/2\pi)E_F/\gamma^2$. The density of states is enhanced near
Dirac point ($E=0$), but as $|E| \rightarrow 0$ it goes zero as
$D(E) \rightarrow |E|\ln|E|$ for both types of screened impurity
scattering. Based on the results of Section III, we expect that
this result is an artifact of the Born approximation, and that the
density of states should in fact be non-zero.  The enhancement of
the density of states can be explained as follows. In normal 2D
system with finite disorder the band edge $E_{{\rm edge},0}$ of
the pure conduction (valence) band is shifted to $E_{\rm edge,imp}
< 0$ ($>0$) and a band tail forms below (above) the band edge of a
pure system. Thus, the density of states in the presence of
impurities is reduced for $E >0$ ($E<0$) because the states have
been shifted by the impurity potential into the band tail.
However, for graphene since the conduction band and the valence
band meet at the Dirac point the band tail (or shift of band edge)
cannot be formed, which gives rise to enhancement of density of
states near Dirac point.

Before concluding we point out that our perturbative calculation
of the graphene density of states assumes that the system remains
homogeneous in the presence of impurities. It is, however,
believed \cite{kn:hwang2006c,rossi} that graphene carriers develop
strong density inhomogeneous (i.e. electron-hole puddle) at low
enough carrier densities in the presence of charged impurities due
to the breakdown of linear scattering. In such an inhomogeneous
low-density regime close to the Dirac point, our homogeneous
perturbative calculation does not apply.

\section{Comparisons to other works}
\label{sec:comparison}

In this section, we compare and contrast our model of disorder and
results to other works in the field.

Peres {\em et al.}\cite{peres2006} studied the effect of disorder
in graphene by considering the effect of vacancies on the
honeycomb lattice.  For a finite density of vacancies, they found
that the density of states at the Dirac point is zero for the
``full Born approximation" (equivalent to our $T$-matrix
approximation) and non-zero for the ``full self-consistent Born
approximation" (equivalent to our self-consistent $T$-matrix
approximation).  Our results are consistent with theirs, even
though the regimes that are studied are different.  Vacancies
correspond to the limit where the impurity potential $U_0
\rightarrow \infty$, whereas this work is more concerned with the
weak impurity-scattering limit.

Pereira {\em at al.}\cite{pereira} considered, among several
different models of disorder, both vacancies and randomness in the
on-site energy of the honeycomb lattice.  They numerically
calculated the density of states for these models of disorder. For
compensated vacancies (same density of vacancies in both
sub-lattices of the honeycomb structure) they found that the
density of states increased around the Dirac point.
(Ref.~\onlinecite{pereira} also studied the case of uncompensated
vacancies, but that has no analogue in our model of disorder.) For
the case of random on-site impurity potential, they find that
``there is a marked increase in the DOS (density of states) at
$E_D$ (the Dirac point)'' and ``the DOS becomes finite at $E_D$
with increasing concentration" of impurities.  Our self-consistent
$T$-matrix approximation result is consistent with their numerical
results, although it should be mentioned again that it strictly
does not apply to the case of vacancies.

Wu {\em et al.}\cite{wu2008} numerically investigated the average
density of states of graphene for the case of on-site disorder.
They found that for weak disorder, the density of states at the
Dirac point increased with both increasing density of impurities
and strength of the disorder. (In their work, they did not absorb
the shift in the band due to the impurity potential in their
definition of the energy, so the Dirac point had a shift $E_D =
xv$ where $x$ is the concentration of impurities and $v$ is the
on-site impurity energy.)  Their numerical results for the minimum
in the average density of states (Fig.~3(b) in
Ref.~\onlinecite{wu2008}) seem to indicate a non-linear dependence
of the value of the minimum as a function of the strength of the
disorder potential, and is at least not inconsistent with
Eq.~(\ref{eq:DoSweak}).

The issue of the effect of screening of the impurity interactions
on the density of states discussed in Section IV, to the best of
our knowledge has not yet been treated in the literature.
Qualitatively, the effect of screened impurities away from the
Dirac point is the increase the density of states, and is
consistent with numerical results for random on-site
disorder.\cite{pereira,wu2008} (Since our treatment of screened
impurities is at the level of the Born approximation, we do not
obtain either a non-zero density of states at the Dirac point nor
resonances in the density of
states.\cite{skrypnyk,pereira,wu2008})

\section{Conclusion}
\label{sec:conclusion}

We have calculated the density of states for disordered graphene.
In the case of unscreened short-ranged impurities, we utilized the
non-self-consistent and self-consistent Born and $T$-matrix
approximations to calculate the self-energy.  Among these, only
the self-consistent $T$-matrix approximation gave a non-zero
density of states at the Dirac point, and the density of states is
a non-analytic function of the impurity potential. We investigated
the density of states in the case of screened short-ranged and
charged impurity potentials at the level of the Born
approximation. We find that, unlike the case of parabolic band
2DEGs, in graphene near the band-edge ({\em i.e.}, the Dirac
point) the density of states is enhanced by impurities instead of
being suppressed. At very low carrier densities, however, graphene
develops strong carrier density inhomogeneity in the presence of
charged impurities, an effect not captured by the homogeneous
many-body theory in our description.

\begin{acknowledgements}
 This work is supported by U.S. ONR, NSF-NRI, and
SWAN.


\end{acknowledgements}

\end{document}